\documentclass[12pt]{article}
\usepackage{epsfig}

     \def\lsim{\raise0.3ex\hbox{$<$\kern-0.75em\raise-1.1ex\hbox{$\sim$}}}
\def\gsim{\raise0.3ex\hbox{$>$\kern-0.75em\raise-1.1ex\hbox{$\sim$}}}
\def\noi{\noindent}

\def\bea{\begin{eqnarray}}  \def\eea{\end{eqnarray}}
\def\beq{\begin{equation}}   \def\eeq{\end{equation}}

\def\beeq{\begin{eqnarray}} \def\eeeq{\end{eqnarray}}

\begin{document}
\begin{center}
{\Large \bf New J/$\psi$ suppression data and the comovers
interaction model} \\

\vskip 8 truemm
{\bf A. Capella$^{\rm a)}$ and D. Sousa$^{\rm b)}$}\\
\vskip 5 truemm

$^{\rm a)}$ Laboratoire de Physique Th\'eorique\footnote{Unit\'e Mixte de
Recherche UMR n$^{\circ}$ 8627 - CNRS}
\\ Universit\'e de Paris XI, B\^atiment 210,
F-91405 Orsay Cedex, France \\

\vskip 5 truemm
$^{\rm b)}$ ECT$^{*}$, Villa Tambosi\\ Strada delle Tabarelle 286, 38050
Villazzano (Trento), Italy
\end{center}

\vskip 1 truecm
\begin{abstract}
New data on the $J/\psi$ suppression both in proton-nucleus and in lead-lead
interactions have been presented recently by the NA50 collaboration. We show
that these data, together with the final ones on sulfur-uranium interactions,
can be described in the framework of the comovers interaction model with a
unique set of three parameters~: the nuclear absorption cross-section, the
comovers interaction cross-section and a single (rescaled) absolute
normalization.
Expectations for $J/\psi$ suppression at RHIC are also discussed.
\end{abstract}

\vskip 1 truecm

\noi LPT Orsay 03-17 \par
\noi March 2003\par
\newpage
\pagestyle{plain}
\section{Introduction}
\hspace*{\parindent}
Before the Quark Matter conference of 2002, the NA50 interpretation of the data
on $J/\psi$ suppression was as follows [1-3]. The $pA$, $SU$ and
peripheral $Pb$
$Pb$ data (up to $E_T \sim 35 \div 40$~GeV) can be described with nuclear
absorption alone, with an absorptive cross-section $\sigma_{abs} = 6.4 \pm
0.8$~mb. At $E_T \sim 40$~GeV there is a sudden onset of anomalous
suppression, followed by
a steady fall off at larger $E_T$.  However, at variance with this
view, the most
peripheral points in $Pb$ $Pb$ collisions lied above the NA50 nuclear
absorption curve -- which
extrapolates $pA$ and $SU$ data. \par

Two important sets of new data have been presented recently \cite{4r}
\cite{5r}. The new NA50 data on $pA$ reactions at 450 GeV/c indicate a
smaller value of $\sigma_{abs}$ than the one given above. However, within
errors, $pA$ and $SU$ data can still be described with a single value of the
absorptive cross-section $\sigma_{abs} = 4.4 \pm 0.5$~mb -- substantially lower
than the previous one \cite{5r}. The new, preliminary, $Pb$ $Pb$ data
\cite{4r}, taken in 2000
with a target under vacuum, are consistent with previous ones except
for the most
peripheral ones -- which are now lower and consistent with the nuclear
absorption curve \cite{4r}. In this way, the NA50 interpretation
remains valid. However, the
new data lend support to the interpretation based on comovers
interaction -- according to which
some anomalous suppression is already present in $SU$ collisions.
Indeed, in the comovers
approach\footnote{For reviews on deconfining and comover interaction models see
\cite{6r}. Alternative models have also been proposed \cite{7r}.} the sudden
onset of anomalous suppression due to deconfinement is replaced by a smooth
anomalous suppression due to comovers interaction. The effect of the
comovers turns out to be
negligibly small in $pA$ but it is sizable in $SU$ interactions. With
the smaller value of
$\sigma_{abs}$ from the new $pA$ data, there is more room for
comovers in $SU$. \par

  The purpose of this work is to study the consistency of the new data
on $pA$ and $Pb$ $Pb$
interactions, together with the final $SU$ ones, with the comovers
interaction model \cite{8r}
\cite{9r}. We proceed as follows. Since the effect of the comovers
suppression is sizable in
$SU$, but negligibly small in $pA$, $\sigma_{abs}$ has to be
determined from the $pA$ data
alone. In previous works [8-9] we have used a value $\sigma_{abs} =
4.5$~mb, as a compromise
between NA38/NA51 \cite{10r} and E537 \cite{11r} $pA$ data. Actually,
it has been shown in
\cite{12r} that the old $pA$ data are also consistent with
$\sigma_{abs} = 4.5$~mb. As
mentioned above, this value has now been confirmed by the recent NA50
data \cite{5r}, and will
be used throughout this paper. The second parameter in the model, the
comovers interaction
cross-section $\sigma_{co}$, can then be determined from the
centrality dependence of the
$J/\psi$ suppression in $Pb$ $Pb$ collisions. Obviously its value is
correlated with that of
the third parameter of the model, the absolute normalization. This
normalization, in
turn, is strictly related to the one in $SU$. The ratio of the $Pb$
$Pb$ to the $SU$
normalizations is equal to $1.051 \pm 0.026$ \cite{5r}. This is a
rescaling factor which takes
into account both isospin and energy corrections. In this way, with
$\sigma_{abs}$ fixed,
the model is strongly constrained.\par

The main drawback of the comovers model [8-9] was precisely a
mismatch of about 30~\% between
the absolute normalizations in $SU$ and $Pb$ $Pb$ \cite{12r}. The
origin of this mismatch is
the following. The high values of the most peripheral $Pb$ $Pb$ data
in the former NA50
analysis required a value $\sigma_{co} = 1$~mb. As stated above, in
the new data
the most peripheral $Pb$ $Pb$ points are substantially lower and require a
lower value of $\sigma_{co} = 0.65$~mb. Indeed, a smaller value of
$\sigma_{co}$ (with
$\sigma_{abs}$ fixed) leads to a flatter centrality dependence of the $J/\psi$
suppression. This change in $\sigma_{co}$ induces a change in the
absolute normalization --
which is now in good agreement with the (rescaled) one obtained in $SU$. \par

The plan of this paper is as follows. In Section 2 we present a short
summary of the comovers
interaction model [8-9]. In Section 3 we apply it to $Pb$ $Pb$
collisions, where the data
allow an accurate determination of both the comovers cross-section
and the absolute
normalization. We also compute the correlation between $E_T$ and
$E_{ZDC}$ -- the
energy of the zero degree calorimeter and discuss the $J/\psi$
suppression in the $E_{ZDC}$
analysis. In Section 4 we show that the $pp$, $pA$ and $SU$ data can
be described using the
same values of $\sigma_{abs}$ and $\sigma_{co}$ as in $Pb$ $Pb$ and a
single (rescaled)
normalization -- obtained from either $SU$ or $Pb$ $Pb$ data. Section
5 contains our
conclusions and expectations for $J/\psi$ suppression at RHIC.

\section{Comovers interaction in the dual parton model}
\hspace*{\parindent}

The cross-section of minimum bias ($MB$), lepton pair ($DY$) and
$J/\psi$ event samples are
given by
  \beq
\label{1e}
I_{MB}^{AB} (b) \propto \sigma_{AB} (b)
\eeq

\beq
\label{2e}
I_{AB}^{DY}(b) \propto \int d^2s \ \sigma_{AB}(b) \ n(b, s)
\eeq

\beq
\label{3e}
I_{AB}^{J/\psi}(b) \propto \int d^2s \ \sigma_{AB}(b) \ n(b,s) \
S_{abs}(b, s) S_{co}(b, s) \ .
\eeq

\noi Here $\sigma_{AB} (b) = \{ 1 - \exp [ - \sigma_{pp} \
AB\ T_{AB}(b)] \}$ where
$T_{AB}(b) = \int d^2s T_A (s) \ T_B(b - s)$, and $T_A(b)$ are profile
functions obtained from the
Woods-Saxon nuclear densities \cite{13r}. Upon integration over $b$ we
obtain the $AB$ total
cross-section, $\sigma_{AB}$. $n(b,s)$ is given by
\beq
\label{4e}
n(b, s) = AB \ \sigma_{pp} \ T_A(s) \ T_B(b - s)/\sigma_{AB}(b) \ .
\eeq

\noi Upon integration over $s$ we obtain the average number of binary
collisions $n(b) = AB \
\sigma_{pp} \ T_{AB}(b)/\sigma_{AB}(b)$. \par

The factors $S_{abs}$ and $S_{co}$ in (\ref{3e}) are the survival
probabilities of the $J/\psi$ due to nuclear absorption and comovers
interaction,
respectively. They are given by \cite{8r} \cite{9r}
\beq
\label{5e}
S^{abs}(b, s) = {[1 - \exp (- AT_A(s) \ \sigma_{abs})] [1 - \exp (- B
\ T_B (b - s) \ \sigma_{abs})]
\over \sigma_{abs}^2 \ AB \ T_A(s) \ T_B(b - s)} \eeq

\beq
\label{6e}
S^{co} (b, s) = \exp \left [ - \sigma_{co} {3 \over 2}
N_{y_{DT}}^{co}(b, s) \ell n \left ( {{3
\over 2} N_{y_{DT}}^{co}(b, s) \over N_f} \right ) \right ]  \eeq

\noi In (\ref{6e}), $N_{y_{DT}}^{co}(b, s)$ is the density of charged
comovers (positives and
negatives) in the rapidity region of the dimuon trigger and $N_f =
(3/\pi R^2_p)(dN/dy)_{y^* =
0} = 1.15$~fm$^{-2}$ \cite{8r,9r,14r}
is the corresponding density in $pp$. The factor 3/2 in (\ref{6e})
takes care of the neutrals.
In the numerical calculations we use
$\sigma_{abs} = 4.5$~mb. The value of
$\sigma_{co}$ and the absolute normalization will be determined from the data.
\par

In order to compute the density of comovers we use the DPM formalism
described in \cite{15r}. It turns out that the
density of charged particles is given by a linear superposition of the
density of participants and
the density of binary collisions with coefficients calculable in DPM.
All details can be found in \cite{8r} and \cite{15r}.\par

Eqs. (\ref{1e}) to (\ref{6e}) allow to compute the impact parameter
distributions of the
$MB$, $DY$ and $J/\psi$ event samples. Experimental results are plotted as a
function of observable quantities such as $E_T$~--
the energy of neutrals deposited in the calorimer. Using the
proportionality between $E_T$ and
multiplicity, we have
\beq
\label{7e}
E_T(b) = {1 \over 2} \ q \ N_{y_{cal}}^{co}(b) \quad .
\eeq

\noi Here the multiplicity of comovers is determined in the rapidity
region of the $E_T$
calorimeter. The factor $1/2$ is introduced because $N^{co}$ is the
charged multiplicity
whereas $E_T$ refers to neutrals. In this way $q$ is close to the
average transverse energy per
particle, but it also depends on the calibration of the calorimeter.
The correlation $E_T - b$
is parametrized in the form \cite{3r,14r}
\beq
\label{8e}
P (E_T, b) = {1 \over \sqrt{2 \pi q a E_T(b)}}
\ \exp \left \{ - (E_T - E_T(b))^2/2 q a E_T(b)
\right \} \ .
\eeq

\noi The $E_T$ distributions of $MB$, $DY$ and $J/\psi$ are then
obtained by folding Eqs.
(\ref{1e})-(\ref{3e}) with $P(E_T, b)$, i.e.
\beq
\label{9e}
I_{AB}^{MB,DY,J/\psi}(E_T) = \int d^2b \ I_{AB}^{MB,DY,J/\psi}(b) \
P(E_T, b) \ . \eeq

\noi The parameters $q$ and $a$ are obtained from a fit of the $E_T$
distribution of the $MB$ event
sample\footnote{Note that the same
value of the parameter $a$ is used in the $MB$, $DY$ and $J/\psi$
event sample. A priori there could
be some differences in the fluctuations for hard and soft processes.
Actually, it has been
claimed in Ref. \cite{8r} that there is a small shift in $E_T$
between minimum bias, on
one hand, and $J/\psi$ or Drell-Yan pair production on the other hand
-- induced by the dimuon
trigger. However, this is of no relevance for the present work,
since, so far, the only 2000
data available have been obtained in the so-called standard analysis
-- in which the
genuine ratio of $J/\psi$ and $DY$ cross-sections is measured.}. Note
that since
$N_{y_{cal}}^{co}(b)$ is nearly proportional to the number of
participants (see Fig. 1 of
\cite{8r}), our fit is practically identical to the one obtained
\cite{14r} using the wounded
nucleon model. Actually, we obtain identical curves to the ones in
Fig. 1 of ref. \cite{3r} --
where the $E_T$ distributions of $MB$ events of 1996 and 1998 are
compared with each other.
The values of the parameters for the 1996 data are $q = 0.62$~GeV and
$a = 0.825$. For the
1998 data, the tail of the $E_T$ distribution is steeper, and we get
$q = 0.62$~GeV and $a =
0.60$ \footnote{At first sight these sets of values look very
different from the ones used by
the NA50 collaboration. Nevertheless, they reproduce the same $E_T$
distribution. This is due
to the fact that the product $qa$, which according to Eq. (\ref{8e}),
determines the width of
the distribution, is very similar in the two cases. As for the
difference in the values of $q$
it is just due to its definition, which is different in the two
approaches (Eq. (\ref{7e}), in
our case).}. In the following we shall use the latter values. Indeed,
according to the NA50
collaboration \cite{2r}, the 1996 data (thick target) at large $E_T$
are contaminated by
rescattering effects -- and only the 1998 data should be used beyond
the knee. \par

The model described above allows to compute the
$E_T$ distribution of $MB$,
$DY$ and $J/\psi$ event samples between peripheral $AB$ collisions
and the knee of the $E_T$
distribution. Beyond it, most models, based on either deconfinement or
comovers interaction, give a
ratio of $J/\psi$ to $DY$ cross-sections which is practically constant -- in
disagreement with NA50 data. A possible
way out was suggested in \cite{9r}. The idea is that, since $E_T$
increases beyond the knee due to
fluctuations, one can expect that this is also the case for the
density of comovers. Since
$N_{y_{DT}}^{co}$ does not contain this fluctuation, it has been
proposed in \cite{9r} to introduce the
following replacement in Eq. (\ref{6e})~:
\beq
\label{10e}
N_{y_{DT}}^{co}(b, s) \to N_{y_{DT}}^{Fco}(b, s) = N_{y_{DT}}^{co}(b, s) \ F(b)
\eeq

\noi where $F(b) = E_T/E_T(b)$. Here $E_T$ is the measured value of
the transverse energy and
$E_T(b)$ is its average value given by Eq. (\ref{7e}) -- which does
not contain the fluctuations. \par

\section{J/$\psi$ suppression in Pb Pb}
\subsection*{a) E$_{\bf T}$ analysis}
\hspace*{\parindent}
The new data \cite{4r} for the ratio of $J/\psi$ over $DY$
cross-sections versus the energy of
the $E_T$ calorimeter are shown in Fig.~1. They are compared with the
results of the comovers
interaction model described in Section 2. As explained there, there
are two free parameters in
the model ($\sigma_{abs} = 4.5$~mb has been fixed)~: the comovers
interaction cross-section
$\sigma_{co}$ (which controls the centrality dependence of the ratio)
and the absolute
normalization. A good description of the data is obtained using
$\sigma_{co} = 0.65$~mb and an
absolute normalization of 47. \par

The only difference between our result and the one in \cite{8r}
resides in the value of
$\sigma_{co}$. Since the effect of the comovers increases with
centrality, a larger (smaller)
value of $\sigma_{co}$ leads to a larger (smaller) variation of the
ratio of $J/\psi$ over
$DY$ cross-sections between peripheral and central collisions. As
mentioned in the
Introduction, in the new NA50 analysis the values of this ratio for
peripheral collisions are
smaller. In order to describe the new data, the value of
$\sigma_{co}$ has to be
reduced. The curve in Fig.~1 corresponds to a
reduction of $\sigma_{co}$ from 1~mb (used in \cite{8r}) to 0.65~mb. \par

Since the values of $\sigma_{co}$ and of the absolute normalization
are correlated, the
decrease of $\sigma_{co}$ induces, in turn, a decrease of the
absolute normalization. While
in \cite{8r} the value of the absolute normalization was about 30~\%
higher \cite{12r} than in
$SU$, the one in Fig.~1 is in good agreement with the $SU$ one. This
will be shown in the next
section.  It is interesting that almost the same value of
$\sigma_{co}$ ($\sigma_{co} =
0.7$~mb) was obtained in \cite{16r} from an analysis of $SU$ data and
former $Pb$ $Pb$ data
\cite{1r} which covered a much smaller centrality range. In
\cite{16r} the absolute
normalizations in $SU$ and $Pb$ $Pb$ were in good agreement with each
other. \par

The $DY$ cross-section in Fig.~1 has been integrated in the dimuon
mass range 2.9 to 4.5~GeV.
Since the $J/\psi$ peak is inside this range, a model is needed in
order to determine the $DY$
cross-section. In the $SU$ analysis, the GRV parton distribution
functions at leading order
(LO) have been used. Therefore, in order to use the (rescaled)
absolute normalization of the
$SU$ data in $Pb$ $Pb$ (or vice versa), the same GRVLO distributions
have to be used in the
latter. This is the case in Fig.~1. In $Pb$ $Pb$ collisions, NA50 has
also analyzed the data
using, instead, MRS 43 distributions. They have found \cite{4r} that
in this case the
absolute normalization is lower by about 10~\%. The comparison of the
comovers model with the
data is presented in Fig.~2. The absolute normalization is 43. The
values of $\sigma_{abs}$ and
$\sigma_{co}$ are, of course, unchanged.

\subsection*{b) E$_{\bf ZDC}$ analysis}
\hspace*{\parindent}
The NA50 collaboration has also measured the $J/\psi$ suppression in
$Pb$ $Pb$ as a function
of the energy of the zero degree calorimeter ($E_{ZDC}$). In Fig.~1,
the results of this
analysis have been plotted as a function of $E_T$, using the measured
correlation between
average values of $E_T$ and $E_{ZDC}$. We see from Fig.~1 that the
data obtained in the two
analysis are consistent with each other, even for very central
events, beyond the knee of the
$E_T$ and $E_{ZDC}$ distributions. This important result has been
predicted in \cite{12r}.
Its physical origin is the following. \par

The energy of the zero degree calorimeter is given by
\beq \label{11e} E_{ZDC}(b) = [A - n_A(b)] E_{in} + \alpha \ n_A(b) \
E_{in} \ . \eeq

\noi Here $n_A(b)$ is the average number of participants at fixed $b$~:
\beq \label{12e}
n_A(b) = A \int d^2s\ T_A(s) \left [ 1 - \exp \left \{ - \sigma_{pp}
\ B \ T_B(b -s) \right
\} \right ] \sigma_{AB}(b) \eeq

\noi $A - n_A(b)$ is the number
of spectator nucleons of $A$ and $E_{in} = 158$~GeV is the beam
energy. While the first term in
the r.h.s. of Eq. (\ref{11e}) gives the bulk of $E_{ZDC}$, the latter
corresponds
to the contamination by secondaries emitted very forward
\cite{17r} -- assumed to be proportional to the number
of participants, $n_A(b)$. Here also the value of $\alpha$ can be precisely
determined from the position of the ``knee'' of the $E_{ZDC}$ distribution
of the $MB$ event sample measured by NA50 \cite{17r}. We obtain $\alpha
= 0.076$ \cite{12r}. \par

Eqs. (\ref{7e}) and (\ref{11e}) give the relation between $b$ and
$E_T$ and $b$ and $E_{ZDC}$,
respectively. These relations refer to average values and do not contain any
information about the tails of the $E_T$ or $E_{ZDC}$ distributions.
Eqs. (\ref{7e}) and (\ref{11e}) also lead to a correlation between
(average values of) $E_T$ and $E_{ZDC}$. This correlation \cite{12r}
gives a good description of the experimental one
\cite{3r}. It is practically a straight line\footnote{This is due to
the fact that
$N_{yca}^{co}(b)$ in Eq. (\protect{\ref{7e}}) is practically
proportional to $n_A(b)$ (see Fig.~1 of \protect{\cite{8r}}).} and
therefore can be accurately extrapolated beyond the knee of the $E_T$
and $E_{ZDC}$ distributions. It turns out that this extrapolation describes
the measured $E_T - E_{ZDC}$ correlation quite well\footnote{One can
understand the physical
origin of this extrapolation if one assumes that a fluctuation in $E_T$ is
essentially due to a fluctuation in $n_A$ -- which, in turn, produces a
corresponding fluctuation in $E_{ZDC}$, via Eq.
(\protect{\ref{11e}}).} -- even for values of
$E_T$ and $E_{ZDC}$ in the tails of the distributions. This result
suggests a correlation
between $b$ and $E_{ZDC}$ of the form \beq
\label{13e}
P(E_{ZDC}, b) = P(E_T, b) \delta \left ( E_T -  E_T(E_{ZDC} \right ) \ .
\eeq

\noi Folding (\ref{1e}) and (\ref{13e}) we
obtain the $E_{ZDC}$ distribution of $MB$ events. It describes
\cite{12r} the one measured
by NA50, not only up to the knee, but also in the tail of the
distribution. This result
shows that the $J/\psi$ supression versus $E_{ZDC}$ is just obtained
from the corresponding one
versus $E_T$ by applying the $E_{ZDC} - E_T$ correlation, even for very central
events beyond the knee of the distributions. In the 1996 and 1998
NA50 data, the $J/\psi$
suppression versus $E_{ZDC}$ indicated some features (``snake
shape'') not present in the
ones versus $E_T$. Such differences are no longer present in the new data.

\section{J/$\psi$ suppression in pA and SU}
\hspace*{\parindent}
Let us  compute next the ratio $R$ of $J/\psi$ over $DY$
cross-sections in $SU$ at 200 GeV/c
per nucleon in our model. We use, of course, the same  values of the
parameters as in $Pb$
$Pb$~: $\sigma_{abs} = 4.5$~mb and $\sigma_{co} = 0.65$~mb. To get
this ratio versus $b$, the
only new ingredient is the multiplicity of comovers -- which is again
computed in DPM, in the
way described in Section 2. To compute
$R(E_T)$, we also need the $E_T - b$  correlation in $SU$, which is
parametrized as in Eq. (\ref{8e}). The parameters $q$ and $a$ have
been obtained from a fit of the $E_T$-distribution of $DY$ given in
\cite{18r}. We obtain $q = 0.69$~GeV and $a = 1.6$. ($R(E_{ZDC})$ has not
been measured in $SU$). In $SU$, data do not extend beyond the knee
of the $E_T$-distribution.
Therefore, effects such as $E_T$ fluctuations, Eq. (\ref{10e}), are
not relevant here.  \par

Our results are shown in Fig.~3. We see that the $E_T$ dependence of
the suppression is reproduced. This indicates that there is, indeed,
room in $SU$ for the
(comparatively small) suppression by comovers. As discussed above
this could also be inferred from the different
central values of $\sigma_{abs}$ obtained in $pA$ and $SU$. The
absolute normalization of
the curve in Fig.~3 is 45. Thus the normalizations in $Pb$ $Pb$ and
$SU$ are consistent with
each other. This normalization is 4~\% smaller than the one obtained
from the $Pb$ $Pb$ data
-- in perfect agreement with the rescaling factor discussed in the
Introduction. \par

In $pA$ collisions, the effect of the comovers is negligible and,
therefore, the
description of the $pA$ data is the same as in the NA50 analysis
\cite{5r}, since, as discussed
above, the value of $\sigma_{abs}$ they obtain is practically
identical to ours. The
corresponding normalization is 20~\% higher than the one we have
obtained in $SU$. This is
also consistent with the expected rescaling factor between the two
systems -- which takes into
account the difference in energy as well as in the rapidity regions
covered by the dimuon
trigger.

\section{Conclusions and outlook}
\hspace*{\parindent}
The NA50 deconfining scenario has been described in the Introduction.
In this work we have
presented a different scenario in which the sudden onset of anomalous
suppression due to
deconfinement is replaced by a smooth one resulting from comovers
interaction. This
anomalous suppression is already present in $SU$ and peripheral $Pb$
$Pb$ collisions. \par

We have presented a description of the NA38/NA50 data on the $J/\psi$
suppression in $pp$,
$pA$, $SU$ and $Pb$ $Pb$ interactions in a comovers model. The model
is strongly
constrained by the existing data. This can be seen in the following
way. In $SU$, the effect of
the comovers is rather small and, thus, once the value of
$\sigma_{abs} = 4.5$~mb is fixed, the
absolute normalization depends little on the exact value of
$\sigma_{co}$. Since the
normalizations in $SU$ and $Pb$ $Pb$ are strictly related, we are
left with a single free
parameter, $\sigma_{co}$, to determine the $J/\psi$ suppression in
$Pb$ $Pb$ (both in absolute
value and centrality dependence). The model is, thus, strongly
constrained and provides a
unified description of the data in the various systems. On the other
hand, it is not
possible to describe the former NA50 data on $Pb$ $Pb$ collisions for
peripheral events in
a consistent way. Indeed, these data require a value $\sigma_{co} =
1$~mb. As shown in
\cite{12r}, this, in turn, leads to a mismatch of about 30~\% between
the absolute
normalizations in $SU$ and $Pb$ $Pb$ systems. Furthermore, with equal
normalizations in
$SU$ and in $Pb$ $Pb$, the $J/\psi$ suppression is always larger in
the latter than in the
former, even for very peripheral events (see Fig. 7 of \cite{12r}).\par

Let us discuss briefly the expectations for $J/\psi$ suppression at
RHIC in the comovers
interaction model. The calculation of the survival probability
$S_{co}$ is quite safe.
Indeed, since $\sigma_{co}$ is a cross-section near threshold, the
same value obtained at SPS
should be used at RHIC. The situation is quite different for
$S_{abs}$. Many authors assume
that $\sigma_{abs}$ is the same at RHIC and at SPS. It has also been
suggested that it can
be significantly larger at RHIC. However, it seems plausible that at
mid-rapidities, nuclear
absorption at RHIC is small due to the fact that, contrary to SPS,
the $c\overline{c}$ pair
is produced outside the colliding nuclei. It is therefore crucial to
have data on $J/\psi$
production in $pA$ interactions at RHIC. If $S_{abs} \sim 1$, the
$J/\psi$ suppression at
RHIC and SPS will be comparable, since the smallness of the nuclear
absorption will be
approximately compensated by the increase of the comovers suppression
-- due to a larger
comovers density at RHIC. Very preliminary data tend to indicate that
this is indeed the
case. Detailed calculations will be presented elsewhere. \par

A quantitative analysis of the new NA50 data in the deconfining
scenario is still missing. On
the other hand, the centrality dependence of the average $p_T$ of
$J/\psi$ is better
described in the comovers approach than in a deconfining scenario
\cite{19r}. At RHIC
energies, a small nuclear absorption in $pA$ collisions (i.e.
$S_{abs} \sim 1$), would be a
very interesting situation in order to discriminate between the
comovers interaction model and
a deconfining scenario. Indeed, in the latter, the shape of the
centrality dependence would be
almost flat for peripheral collisions (below the deconfining
threshold) and would decrease
above the threshold. Such a behavior would be a clear signal of
deconfinement. On the
contrary, in the comovers scenario, the fall-off would be continuous,
from peripheral to
central collisions, and determined by the same value of $\sigma_{co}$
obtained from CERN
SPS data.  \par \vskip 5 truemm

\noi {\large \bf Acknowledgements} \par
We thank N. Armesto for discussions and M. Gonin, L. Kluberg and E.
Scomparin for information
on the NA50 data.

\newpage \section*{Figure Captions}

\noi {\bf Fig. 1 :} Ratio of $J/\psi$ to $DY$ cross-sections versus
$E_T$ in $Pb$ $Pb$ collisions at 158 GeV/c per nucleon (solid
line).  The preliminary data are from
\cite{4r}. GRVLO parton distribution functions have been used in
order to calculate the $DY$
cross-section in the mass range 2.9 to 4.5~GeV.\\

\noi {\bf Fig. 2 :} Same as in Fig. 1, using MRS 43 parton
distribution functions.\\

\noi {\bf Fig. 3 :} The ratio of $J/\psi$ to $DY$ cross-sections as a
function of $E_T$ in
$SU$ collisions at 200 GeV/c per nucleon.
The data are from \cite{18r}.\\

\newpage

\centerline{{\epsfysize12cm \epsfbox{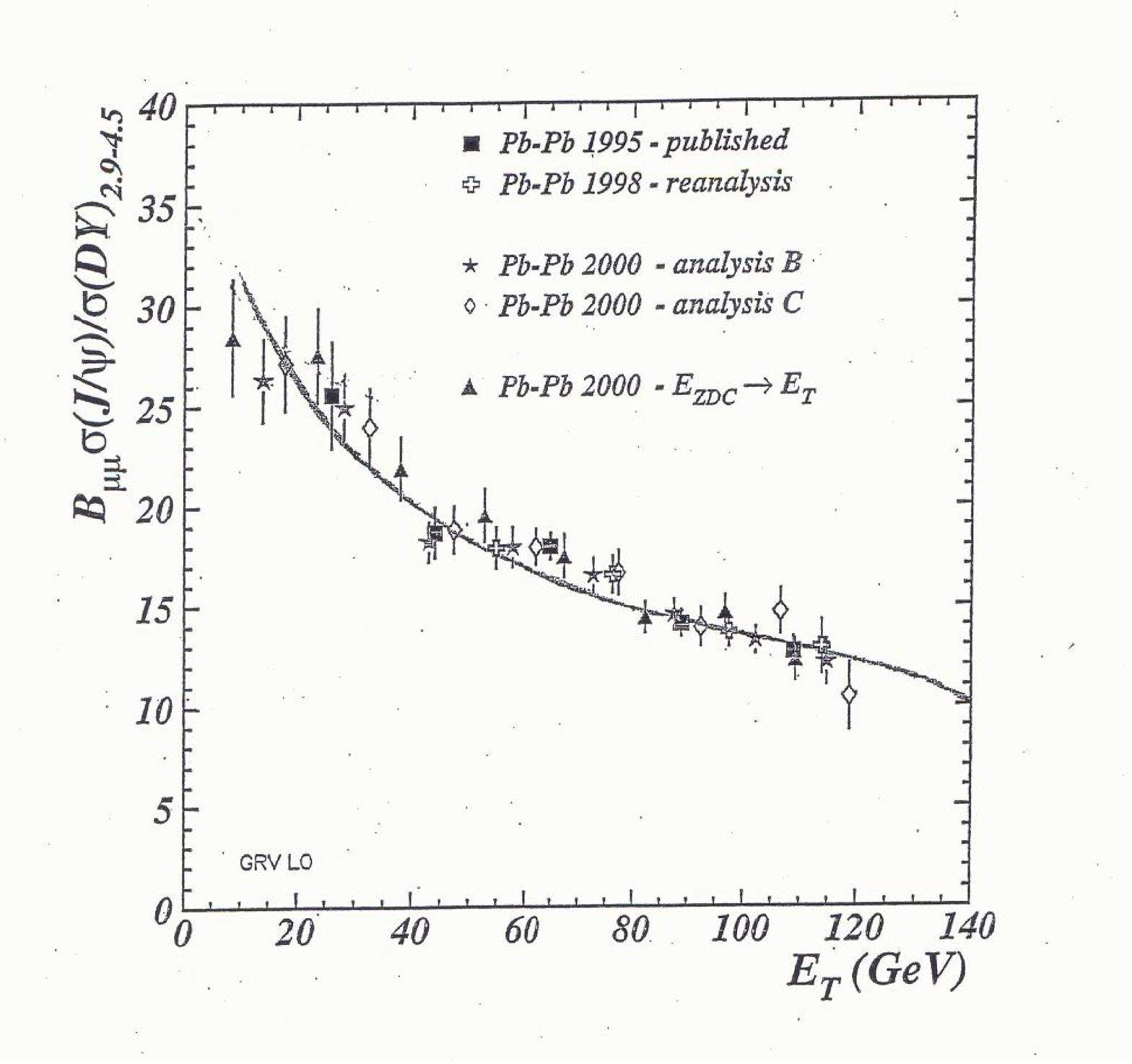}}}
\centerline{Figure 1}

\centerline{{\epsfysize12cm \epsfbox{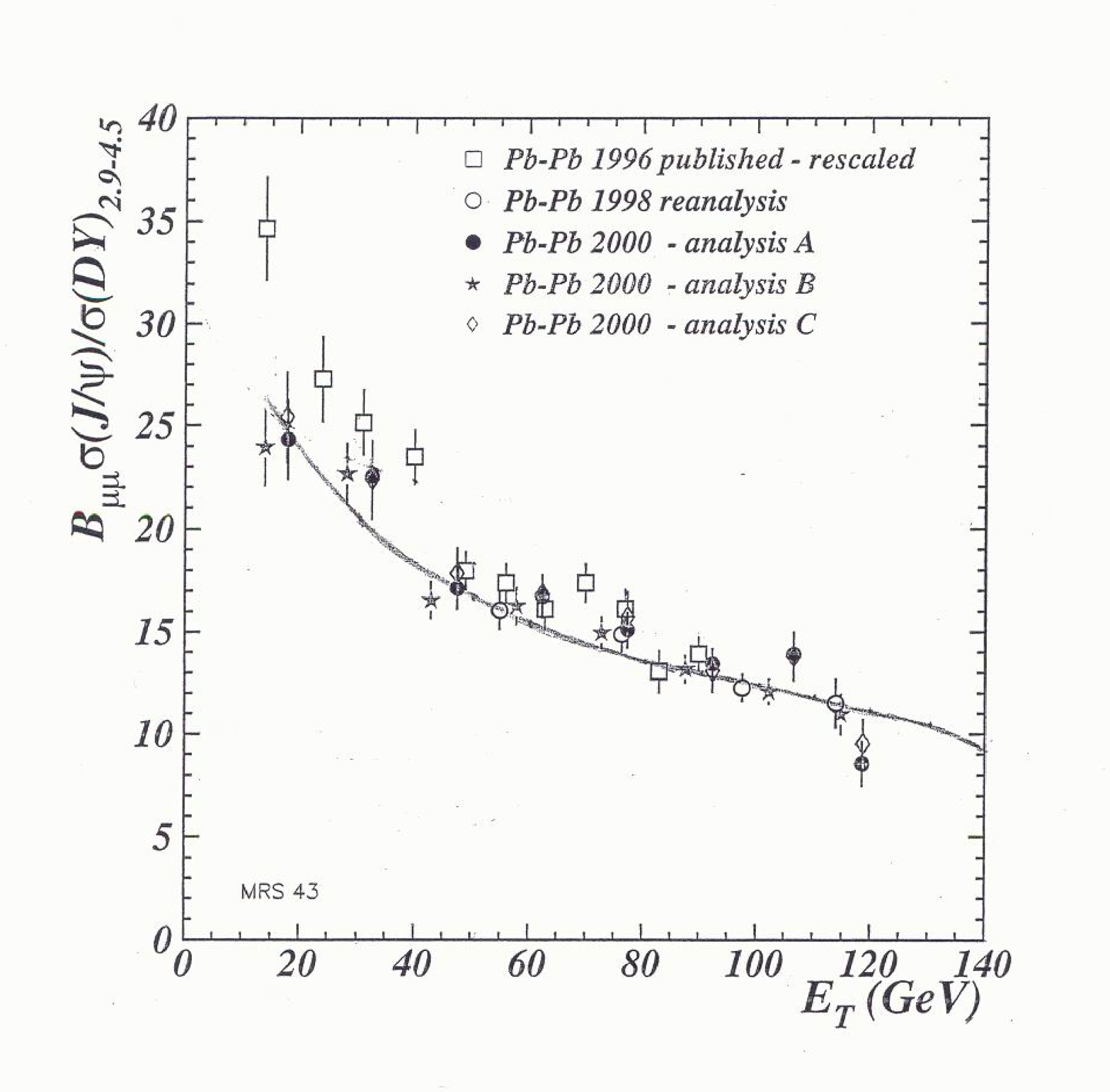}}}
\centerline{Figure 2}

\centerline{{\epsfysize8cm \epsfbox{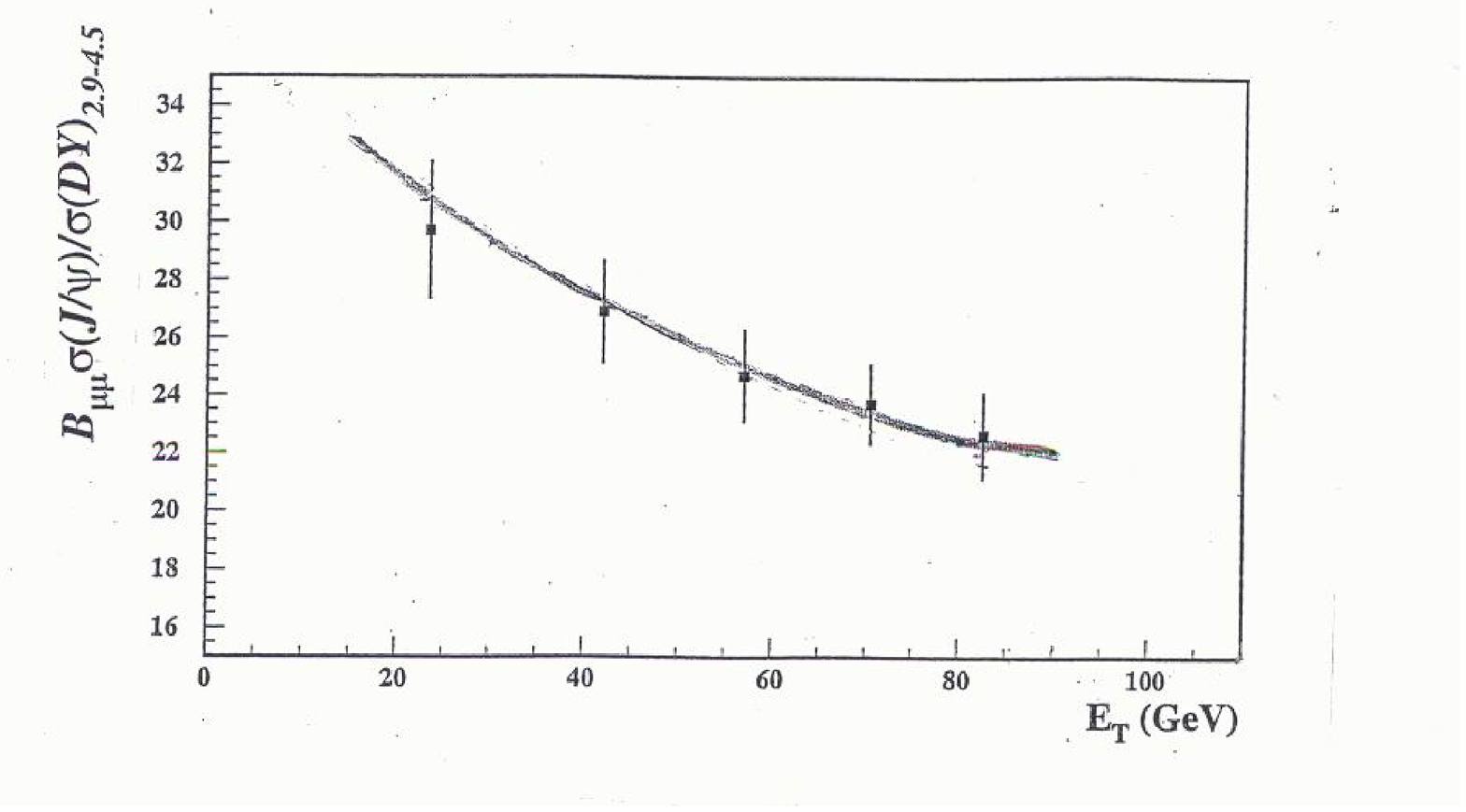}}}
\centerline{Figure 3}


\newpage
  \begin{thebibliography}{99}

\bibitem{1r} NA50 collaboration, M. C. Abreu et al, Phys. Lett. {\bf
B410}, 327 (1997); ibid
page 337.

\bibitem{2r} NA50 collaboration, M. C. Abreu et al., Phys. Lett. {\bf
B450}, 456 (1999).
\bibitem{3r} NA50 collaboration, M. C. Abreu et al., Phys. Lett. {\bf
B477}, 28 (2000).
\bibitem{4r} NA50 collaboration, L. Ramello in Proceedings Quark
Matter 2002, Nantes
(France), 18-24 July 2002 (to be published).

  \bibitem{5r} NA50 collaboration, P. Cortese in Proceedings Quark
Matter 2002, ibid. \\
NA50 collaboration, R. Sahoyan in Proceedings XXVII Rencontres de
Moriond, Les Arcs (France),
March 2002.

\bibitem{6r} C. Gerschel and J. Hufner, Ann. Rev. Nucl. Part. Sci.
{\bf 49}, 255 (1999). \\
R. Vogt, Phys. Rep. {\bf 310}, 197 (1999).\\
H. Satz, Rep. Prog. Phys. {\bf 63}, 1511 (2000).

\bibitem{7r} J. Qiu, J. P. Vary, X. Zhang, hep-ph/9809442; \\
A. K. Chandhuri, preprint Variable Energy Cyclotron Center, Calcutta,
India (September 14, 2001).

\bibitem{8r} A. Capella, A. Kaidalov and D. Sousa, Phys. Rev. {\bf
C65}, 054908 (2002).
\bibitem{9r} A. Capella, E. G. Ferreiro and A. Kaidalov, Phys. Rev.
Lett. {\bf 85}, 2080 (2000).\\
N. Armesto et al., Nucl. Phys. {\bf A698}, 583 (2002).

\bibitem{10r} NA38 collaboration, M. C. Abreu et al, Phys. Lett. {\bf
B466}, 408 (1999);\\
NA51 collaboration, M. C. Abreu et al., Phys. Lett. {\bf B438}, 35 (1998).

\bibitem{11r} FNAL E537 collaboration, M. Leitch et al., Phys. Rev. Lett.
{\bf 84}, 3256 (2000).

\bibitem{12r} A. Capella and D. Sousa, nucl-th/0110072, unpublished.

\bibitem{13r} C. W. Jager et al., Atomic Data and Nuclear Tables {\bf
14}, 485 (1974).

\bibitem{14r} D. Kharzeev, C. Louren\c co, M. Nardi and H. Satz, Z. Phys. {\bf
C74}, 307 (1997).

\bibitem{15r} A. Capella and D. Sousa, Phys. Lett. {\bf B511}, 185 (2001).

\bibitem{16r} N. Armesto and A. Capella, Phys. Lett. {\bf B430}, 23 (1998).

\bibitem{17r} NA50 collaboration, M. C. Abreu et al., Phys. Lett. {\bf
B521}, 195 (2001).

\bibitem{18r} NA38 collaboration, M. C. Abreu et al., Phys. Lett. {\bf
B449}, 128 (1999).

\bibitem{19r} N. Armesto, A. Capella and E. G. Ferreiro, Phys. Rev.
{\bf C59}, 395
(1999).\\
A. K. Chaudhuri, nucl-th/0212046.
  \end{thebibliography}
\end{document}